# The Age of Old Magellanic Cloud Clusters. I: NGC 2257

V. Testa[1], F.R. Ferraro[2], E. Brocato[3], V. Castellani[3,4]

[1] *Dipartimento di Astronomia, Università di Bologna, Bologna, Italy*

[2] *Osservatorio Astronomico, Bologna, Italy*

[3] *Osservatorio Astronomico di Collurania, Teramo, Italy*

[4] *Dipartimento di Fisica, Università degli Studi, Pisa, Italy*

**ABSTRACT**

Deep CCD photometry down to $V \sim 24.5$ is presented for the old globular cluster NGC 2257 in the Large Magellanic Cloud. The main results of this study can be summarized as follows: i) the Red Giant Branch (RGB) Bump has been detected at $V = 18.7 \pm 0.1$, and it is compatible with determination of this feature in Galactic Globular Clusters; ii) using the currently accepted methods of dating Globular Clusters, NGC 2257 seems to be 2-3 Gyrs younger than a typical Galactic Globular; iii) the Helium abundance has been estimated to be $Y = 0.21 \pm 0.03$ using the R-method as calibrated by Buzzoni et al. (1983).

## 1 INTRODUCTION

This is the first of a series of papers devoted to investigating the ages of old globular clusters in the Magellanic Clouds (hereafter MC), in order to allow a comparison with the Galactic Globular Clusters (GGCs) system.

The age of stellar systems is one of the most important astrophysical parameters. As is well known, the galactic halo is mainly populated by old clusters, whose ages should provide not only fundamental information on the early evolution of the Galaxy, but also severe constraints on the age of the Universe. This is not the case for the MC, where one finds globulars as young as 10 million years, if not lower (Barbero et al. 1990). However, the MC do have some globular clusters whose characteristics closely resemble those of GGCs, with CMD morphologies quite similar to that typical of their galactic counterparts. The determination of age and abundances for these MC clusters will provide fundamental information about the very early history of the MC system and about possible interactions between the MC and the Galaxy, showing in particular whether or not the MC started making clusters at the same cosmic era and in similar initial conditions.

Among these MC old clusters, we choose as a first target NGC 2257, which appears of particular interest since according to Stryker 1983, Nemec Hesser and Ugarte 1985 (NHU) and Walker 1989 (W89) it is probably the oldest MC cluster. NGC 2257 lies $9^o$ NE to the LMC bar, in a poorly field-contamined area and it has been recognized as intermediate Oosterhof type (NHU). Available metallicity estimates, as discussed by W89, give [Fe/H] = $-1.8 \pm 0.1$ and indications that the reddening E(B-V) is not greater than 0.04 are provided by the same author.

Photometric results available in the literature concerning either the bright stellar population of the cluster or accurate determination of magnitude and periods for cluster RR Lyrae (see W89 and references therein), already disclose that NGC 2257 shows a marked similarities with typical galactic globulars of intermediate metallicity like NGC 5897 and M 3. Stryker (1983) presented photometry results down to $V \approx 22$, finding indication for the turn off (TO). However, we still lack of investigation clearly detecting the TO luminosity which is the only reliable way to obtain indication on the cluster age.

In this paper we present B,V photometry down to $V \sim 24.5$, i.e. about 2 mag below the TO point, with the aim of estimating the age of the cluster on the basis of the various procedures currently available. The observations and reductions are described in Sect. 2. The overall properties of the CMD are discussed in Sect. 3. In Sect. 4 we derive an estimate for the distance of NGC 2257. Sect. 5 is devoted to some comparisons with NGC 5897. In Sect. 6 we discuss the detection of the Turnoff (TO) region and the age of the cluster. Conclusions are presented in Sect.7 .

## 2   OBSERVATION AND REDUCTIONS

### 2.1   Observations

A set of B,V CCD frames was secured during one photometric night in Dec. 1990 using the NTT 3.5m ESO telescope, equipped with EMMI. The chip was a Tektronix 1024x1024 px, 0.44 arcsec/px sized. The cluster was centered in the NE quarter of the CCD frame in order to have, in the same frame, a field sky region far enough from the cluster center to obtain information on the field population. We obtained short exposures (V 15s, B 30s) to study the bright portion of the CMD, intermediate exposures (V 120s, B 360s) for the region between the Horizontal Branch (HB) and the upper Subgiant Branch (SGB), and deep exposures (V 300s, B 900s) to reach the Main Sequence (MS). All the frames were secured during good seeing condition (0.8″ FWHM). Because of the crowding conditions, we cut off a circle of 60 px radius around the cluster center on the intermediate frames, and one of 110 px on the deep ones, while the bright portion was measured up to the cluster centre. All the frames were referred to the master intermediate one (V 120s), where the cluster centre is located at the coordinates $X_c = 851, Y_c = 816\ px$. The X and Y coordinates are in pixels starting from the lower left (SW) corner of the field (see Figure 1).

Bias exposure has been obtained from the median average of 9 individual frames. Flat-field exposures were taken at the beginning and the end of the night in order to correct the detector response to uniform sensitivity. The level of flat field exposures was 15000-20000 ADU. Both B and V

Figure 1. Map of the observed field in NGC 2257. Some stars have been identified with the number they have in Tables 1a,b.

Table 2. Mean Ridge Lines for NGC 2257.

| RGB | | 0.690 | 20.00 | 0.478 | 21.82 |
|---|---|---|---|---|---|
| 1.770 | 16.30 | 0.680 | 20.10 | 0.436 | 21.88 |
| 1.620 | 16.36 | 0.670 | 20.20 | 0.391 | 22.00 |
| 1.460 | 16.42 | 0.670 | 20.30 | 0.372 | 22.10 |
| 1.290 | 16.49 | 0.670 | 20.40 | 0.358 | 22.20 |
| 1.200 | 16.59 | 0.660 | 20.50 | 0.351 | 22.30 |
| 1.120 | 16.78 | 0.670 | 20.40 | 0.348 | 22.40 |
| 1.050 | 17.03 | 0.660 | 20.50 | 0.350 | 22.50 |
| 0.960 | 17.54 | 0.650 | 20.60 | 0.355 | 22.60 |
| 0.870 | 18.14 | 0.650 | 20.70 | 0.360 | 22.70 |
| 0.800 | 18.73 | 0.650 | 20.80 | 0.365 | 22.80 |
| 0.770 | 19.10 | 0.640 | 20.90 | 0.370 | 22.90 |
| 0.760 | 19.20 | 0.630 | 21.00 | 0.389 | 23.00 |
| 0.750 | 19.30 | 0.620 | 21.18 | 0.475 | 23.82 |
| 0.740 | 19.40 | 0.600 | 21.41 | **HB** | |
| 0.730 | 19.50 | 0.580 | 21.55 | -0.015 | 19.42 |
| 0.720 | 19.60 | 0.560 | 21.62 | 0.056 | 19.15 |
| 0.710 | 19.70 | **SGB–MS** | | 0.140 | 19.07 |
| 0.710 | 19.80 | 0.538 | 21.72 | 0.319 | 18.98 |
| 0.700 | 19.90 | 0.526 | 21.76 | 0.531 | 18.89 |

frames have been flattened to 0.2% peak to peak.

### 2.2 Magnitudes, colours, and positions

The adopted reduction procedure has been widely reviewed in Ferraro *et al.* (1990). According to this procedure the reduction package ROMAFOT (Buonanno *et al.* 1979, 1983), in the version optimized for undersampled images (Buonanno and Iannicola 1988) was used to detect stars on each frame, to transform the instrumental magnitudes and coordinates to the frame assumed as reference and, finally, to get an homogeneous set of CCD instrumental magnitudes, colours and positions.

Since a reliable sequence of photoelectric primary standards was not available for that night, we used a set of $\sim 350$ stars in common with W89 in order to calibrate our data.

The following relations have been obtained:

$$B_W = b - 0.05(b - v) + 30.35$$

$$V_W = v + 30.41$$

where $B_W$ and $V_W$ are the magnitudes listed by W89 and $b$ and $v$ are our instrumental magnitudes.

In Table 1a we list final magnitudes, colours and positions for the 3169 non variable stars in the cluster region ($R < 400px$ : 176"), whereas Table 1b reports the final magnitudes, colours and coordinates for the 1787 stars in the outer regions ($R > 400px$). We present in Figure 1 a computer map of some reference stars from the list of Table 1. Colour Magnitude Diagrams (CMD) obtained from stars listed in Tables 1a and 1b are shown in Figure 2a,b respectively. These CMDs are assumed as representative of the cluster and the field population.

## 2.3 Photometric errors and field contamination

The transformations of instrumental magnitudes at various exposure times to an homogeneus scale offers the possibility to compute the frame to frame scatter of magnitudes for each individual star. Thus the internal accuracy of our CCD measurements can be estimated from the *rms* frame-to-frame scatter of the instrumental magnitudes of each individual star computed according to the formula given in Ferraro *et al.* (1991), where the weights have been computed taking into account the quality of the CCD frame, seeing conditions etc.

The rms-values ($\sigma$) in each 0.5-mag bin, for the two filters, are plotted in Figure 3a,b respectively as a function of the final mean magnitude. Only errors for stars brighter than $V > 17$ and $B > 17$ have been plotted since the estimate of the errors for the brightest stars, measured almost in the central zone, where only one frame per filter was measured, has been made using the scatter of the magnitudes obtained during the simulation of artificial stars to determine the completeness factor, obtaining $\sigma \leq 0.01$ in both filters. However, it is important to point out that the final errors include the combination of two more contributions: the intrinsic error involved in the calibration procedure and the uncertainties in the various transformations.

**Figure 2.** (a) - Cluster CMD; the detected stars listed in Table 1a lying at $R \leq 400px$ from the cluster centre; the horizontal line separates the bright amd the faint samples; (b) - CMD of the field: stars listed in Table 1b all stars lying at $R > 400px$ form the cluster centre.

**Figure 3.** Internal photometric errors of the measurements; the mean values of the frame-to-frame scatter computed in bins of 0.5 mags. are plotted versus V (panel (a)) and B (panel (b)) final magnitude.

**Figure 4.** Colour distribution of the stars in Tabel 1a in the range $22.0 < V \leq 22.5$.

**Table 3.** Contaminating Field Stars per square arcmin.

|  | $B-V < 0$ | $0 \leq B-V < 1$ | $B-V \geq 1$ |
|---|---|---|---|
| $V \leq 18.0$ | 0.0 | 0.6 | 0.2 |
| $18.0 \leq V \leq 20.5$ | 0.0 | 2.0 | 0.7 |
| $20.5 \leq V \leq 22.0$ | 0.0 | 5.2 | 1.0 |
| $V \geq 22.0$ | 2.0 | 36.4 | 5.0 |

A preliminary inspection of the CMDs presented in Figure 2 reveals that the field contamination seems not to affect significantly the definition of the ridge lines of the main branches, at least at the brighter luminosity. At the lower luminosity, one finds that - for each given interval of luminosity - the distribution of star colours shows a steep maximum around a quite well defined colour values, which can be fairly assumed as ridge line. Figure 4 shows an example of the procedure followed, relatively to the magnitude bin $22.0 < V \leq 22.5$.

Table 2 presents the values we will adopt as normal points for the various branches. The contamination from background and foreground stars can be evaluated dividing the CMD into cells and computing the number of field stars expected in each cell, using the CMD of portion of the frame assumed as field and scaling the area of the field to that of the cluster. In Table 3 we report the expected number of contaminating objects, per square arc min in each interval of magnitude and colour, in the cluster area.

### 2.4 Completeness

To test the completeness achieved in our measures following the procedure described in Ferraro *et al.* (1990), we randomly added to the original frames a set of stars ($N_{sim}$) of known magnitude and colours, and re-reduced each frame.

**Figure 5.** Completeness curves for the inner region (panel (a), $R < 60px$) and the intermediate region (panel (b), $60 \leq R < 110px$).

The output of the procedure is the ratio of the number of stars detected ($N_{out}$) with respect to $N_{sim}$.

In order to test the dependence of the completeness level from the crowing conditions, we performed the procedure in three annuli at different distance from the cluster center, up to $V = 20.3$:

*annulus (a)* - $r < 26.4''$  *i.e* : $r < 60px$

*annulus (b)* - $26.4'' < r < 66''$  *i.e.* : $60px < r < 150px$

*annulus (c)* - $r > 66''$  *i.e.* : $r > 150px$.

Figure 5a,b shows the ratio $\Lambda = N_{out}/N_{sim}$ as a function of the V magnitude in the (a) and (b) annuli obtained from the whole procedure. Star detection losses increase at fainter magnitudes, but the completeness degree is always greater than 85 % at $V < 19.5$. Annulus (c) is almost complete ($\Lambda \sim 1$) even at the fainter magnitudes.

However, no completeness estimate has been made below $V = 20.3$, because, at faint magnitudes, we were interested only to the determination of the mean ridge lines and for this reason we considered only a selected sample of stars with the best photometry. The above procedure yields only "first order" corrections (Mateo 1988), but we believe that the residual incompleteness still affecting our samples should not bias our conclusions.

## 3 THE COLOUR MAGNITUDE DIAGRAM

### 3.1 The CMD overall morphology

The CMD plotted in Figure 2 shows that the photometry presented here reaches $V \sim 24.5$, increasing by about two magnitudes the previous photometries. Our CCD frame covered a large region of the cluster so that we are confident we sampled a large fraction of the light of the cluster making the presented CMD a reliable "picture" of the star population in NGC 2257.

The main characteristics of the diagram can be summarized as follows:

1. The giant branch appears fairly well defined and populated also in the very upper part;

2. The HB is populated mainly in the blue side of the instability strip, with the possible occurrence of few very hot stars at its blue end;

3. The region of the TO and the upper portion of the Main Sequence is defined and well populated at least $\sim 2$ mag below the cluster TO.

4. A puzzling features is that, below $V \simeq 17$ the giant branch becomes steep, as expected in metal poor clusters, whereas the brighter portion of the giant branch flattens as in typical metal rich clusters.

The field CMD (fig. 2b) does not allow one to simply mark the flattened stars as no cluster members. We note that the star 895 corresponds to star 491 in W89 (V=15.74, B-V=1.61), differing from our measurement by $\Delta V = 0.49, \Delta(B - V) = -0.18$. This may be and indication of variability, but no firm conclusions can be drawn. Apart from star 895, the flattening of the RGB is present also in Walker's diagram, and S83 considers the possibility that the reddest stars may be incorrectly measured or be variables (see refs. in S83 and followings sections). It may be noted that on our frames there is no sign of saturation, nor of severe crowding affecting these stars.

Certainly, small number statistics may also play a substantial role and the lack of very red standard stars for calibration may be responsible of the unexpected flattening. However, we cannot provide a definitive explanation on this change of slope of the RGB, further observational investigation are the required using near IR filters or spectroscopic measurements.

### 3.2 The Giant Branch and metal abundance indicators

We can measure suitable quantities on the RGB in order to get a photometric estimate of the parameters $\Delta V$ (Sandage and Wallerstein 1960), $S$ (Hartwick 1968) and $(B-V)_{o,g}$ (Sandage and Smith 1966), and use their calibration in terms of [Fe/H] to obtain indication of the metallicity of the cluster. A summary of these relations can be found in Ferraro, Fusi Pecci, Buonanno (1992, hereafter FFB) (see their Table 5). In order to determine these parameters let us adopt a figure for the interstellar reddening $E(B - V)$ and the magnitude level of the zero age horizontal branch $V_{HB}$.

**Figure 6.** Observed RGB luminosity function: Panel (a) shows the differential LF, the arrow indicates the location of the RGB–Bump; panel (b) shows the integral LF, the dashed lines are the linear fits to the regions above and below the RGB-Bump.

All the available estimates of the reddening in NGC 2257 (listed by W89) indicate a low value. W89 finally adopted $E(B - V) = 0.04$ though some methods suggested an even lower value. In the following we will adopt $E(B - V) = 0.04$.

The measure of $V_{HB}$ in NGC 2257 can be obtained by looking at the histogram of the star counts in the HB *vs* the V magnitude. We assume as magnitude of the ZAHB, the magnitude of the lower bound of the histogram's "bell" of the HB ($0.20 < B - V < 0.60$). This yields $V_{HB} = 19.10 \pm$ 0.05, which is in good agreement with the value obtained by W89 who found $V_{HB} = 19.12 \pm 0.05$ (from $V_{RR} = 19.03 \pm 0.05$ and considering that the difference between the mean magnitude of the RR Lyrae and the ZAHB level is 0.09 mag). Using this determination and assuming $E(B-V) = 0.04$ we derive

$$\Delta V = 2.6 \pm 0.1$$

$$(B-V)_{o,g} = 0.73 \pm 0.01$$

$$S = 5.9 \pm 0.1$$

Adopting the relations listed in Table 5 of FFB we obtain the mean values of metallicity for each observable:

$$[Fe/H]_{\Delta V} = -1.48 \pm 0.05$$

$$[Fe/H]_{(B-V)_{o,g}} = -1.79 \pm 0.04$$

$$[Fe/H]_S = -1.72 \pm 0.06.$$

It is interesting to note that, as expected, the estimate obtained with $\Delta V$ method disagrees with the two others; this is the consequence of the flattening of the RGB in its bright portion, but we cannot claim a high metal content in the cluster to justify this result. W89 states that these stars are subluminous for a metal-poor giant branch, so that the membership to the cluster is questionable and no firm conclusion can be established. Moreover, Olszewski *et al.* 1991 made spectroscopic measurements of two luminous giants, namely stars LE7 (our 862) and LE8 (our 866), to get an indication on the metallicity and velocity of the cluster. They find that the two stars are relatively more metal-rich than the cluster ($[Fe/H] \sim -0.4$). On the other hand, the five reddest stars could be variable AGB stars, according to Frogel and Elias 1988 (their Fig. 1), who discuss the red variables in GCs, concluding that, at intermediate metal poor abundances, the red variables have bolometric magnitudes slightly less luminous than the RGB tip; in this case NGC 2257 should present a $V_{tip}$ brighter than what measured on our CMD, and so the value of $\Delta V$ increase to about 2.9,3.0 thus giving a metal abundance calibration compatible with the other two calibrations, and with the value found by W89. More data are necessary to establish firm conclusions.

Since stars at the RGB tip are at least less reliable (low statistic, suspected variability, maybe field stars) than the lower part of the RGB we prefer $[Fe/H] = -1.7$ as "photometric" measure of the cluster metal abundance negleting the $\Delta V$ method. The errors we associate with the metallicity estimations are formal errors since some photometric estimates of the cluster metallicity are, of course, sensible functions of the assumed reddening. Would E(B-V) be decreased to 0.00, we had $\Delta[Fe/H]_{(B-V)_{o,g}} = 0.2$, $\Delta[Fe/H]_{\Delta V} = 0.07.$, while $[Fe/H]_S$ is formally independent on the reddening but is heavily affected by other uncertainties in the photometry (as, for example, a uncorrect application of the colour eq., etc). For this reason we adopt a conservative value of $\pm 0.2$ for the error in the metallicity. In the following we will adopt [Fe/H]$= -1.7 \pm 0.2$.

### 3.3 The RGB Bump

In order to study the Luminosity Function of the RGB down to the SGB we selected stars in the magnitude range $16.0 < V < 20.3$ lying $\pm \sigma$ from the cluster fiducial line listed in Table 2 ($\sigma$ ranging from 0.05 up to 0.2 at different magnitude level).

In Figure 6a,b the RGB LFs, differential and integral, have been plotted as a function of the V magnitude, in order to detect the existence of the RGB–bump. As emphasized by Fusi Pecci *et al.* 1990 the differential and integral LFs offer two independent tools to identify the RGB-Bump. The plot shows this feature, visible in both LFs, located at $V = 18.7 \pm 0.1$. This feature marks the evolutionary stage at which the H–burning shell passes through the chemical discontinuity left by the maximum penetration of the convective envelope (see Renzini and Fusi Pecci 1988 and refs. therein). The RGB–bump has been identified in several GGCs (Fusi Pecci *et al.* 1990, Ferraro *et al.* 1994a), thus confirming the theoretical expectations, though with the known discrepancy in the exact zero point of the relation.

Coupled with the HB magnitude, $V_{HB}$, the value of $V_{bump}$ can be used to compute the parameter $\Delta V_{bump}^{HB} = V_{bump} - V_{HB}$ defined by Fusi Pecci *et al.* 1990. The value we obtain for NGC 2257 is $\Delta V_{bump}^{HB} = -0.40 \pm 0.12$. Figure 7 shows $\Delta V_{bump}^{HB}$ as a function of the hyperbolic arcsine of the metallicity for all the clusters for which this parameter has been determined. Open circles refer to the data listed by Fusi Pecci *et al.* 1990 and Ferraro *et al.* 1994a for Galactic Globulars, and the solid line is a best fit for those data,

$$\Delta V_{bump}^{HB} = 0.33 sinh^{-1}\left(\frac{Z}{0.00025}\right) - 0.68.$$

As can be seen, the result for NGC 2257 (the filled circle) fits the relation well.

### 3.4 HB morphology and Helium abundance

The HB-morphology of NGC 2257 shows a concentration on the blue side, but it is not as extended in the blue as other GGC with similar metallicity. In order to quantitatively describe the behaviour of NGC 2257 we adopt the parameter defined in Lee (1990) ($LZ = (B-R)/(B+V+R)$) where B and R are the HB stars bluer and redder than the instability strip respectively and V the variables, and compare it with other Galactic Globulars. The counts have been corrected for the completeness factor obtained as described above, and the value we obtain is $LZ = 0.47$, typical of a normally "blue" HB. In order to determine the number of stars – $N_{AGB}$, $N_{RGB}$, $N_{HB}$ – to insert into the ratios, we followed the assumptions made by Buzzoni *et al.* (1983).

**Figure 7.** Observed values of $\Delta_{Bump}^{HB}$ as a function of $sinh^{-1}(Z/0.00025)$. Open circles are from Fusi Pecci *et al.* 1990, Ferraro *et al.* 1994a and the filled circle is NGC 2257. The solid line represents the relation found by Fusi Pecci *et al.* 1990.

In Table 4 we give the number of stars we counted in each branch. A small correction for incompleteness was applied, and the five reddest stars ($(B-V) > 1.5$) close to the RGB tip were attributed to the RGB. The ratios computed on the basis of these populations are in quite good agreement with the results found for GGCs (see Buzzoni *et al* 1983, Ferraro 1992a), $R1 = N_{AGB}/N_{RGB} = 0.16 \pm 0.05$,

**Table 4.** Star Counts in the various Branches.

| RGB | blue HB | RR HB | red HB | AGB |
|---|---|---|---|---|
| 134 | 101 | 37 | 25 | 21 |

$R2 = N_{AGB}/N_{HB} = 0.13 \pm 0.04$, $R = N_{HB}/N_{RGB} = 1.22 \pm 0.21$, $R' = N_{HB}/(N_{AGB} + N_{RGB}) = 1.05 \pm 0.17$. The last two ratios can be used to estimate the primordial helium abundance using equation 11 and 13 by Buzzoni *et al* 1983. From the quoted values we get $Y_R = 0.21 \pm 0.03$ and $Y_{R'} = 0.21 \pm 0.03$. These values are slighlty greater than those found by W89 ($Y = 0.19 \pm 0.02$). As stated in Section 3.2 the five reddest stars could be AGB stars, considering this possibility we found $R1 = 0.20 \pm 0.06$, $R2 = 0.16 \pm 0.04$, $R = 1.26 \pm 0.21$ and $R' = 1.05 \pm 0.17$, values still quite compatible with what found in GGCs.

## 4   THE DISTANCE OF NGC 2257

As well known, one can use various different scenarios to get the absolute magnitude of the HB at the colour interval of the RR Lyrae variables, depending on the zero-point and slope of the adopted $M_V^{HB} vs.[Fe/H]$ – calibration (see for a discussion Buonanno, Corsi and Fusi Pecci 1989, hereafter BCF) The absolute magnitude of the HB can vary from $M_V^{RR} = 0.48$ using the relation obtained by Walker (1992):

$M_V^{RR} = 0.15[Fe/H] + 0.73$

up to $M_V^{RR} = 0.74$ at $[Fe/H] = -1.6$ using the recent results by Layden, Hanson and Hawley (1994). In the following we assume the mean value between these two extreme values $M_V^{RR} = 0.61 \pm 0.13$. The differences in these two values are indicative of the uncertainties that still affect the determinations of the distances of GGC.

Considering that $V^{HB} = 19.10$ and $E(B - V) = 0.04$ we obtain a distance modulus $(m - M)_V = 18.4 \pm 0.2$. This value rises to $(m - M)_V = 18.5 \pm 0.2$ if we consider $E(B - V) = 0.0$. From these values a mean distance of $d_\odot = 50 \pm 10 Kpc$ has been assumed for NGC 2257.

## 5   A TWIN-METALLICITY GGC: NGC 5897

The shape of the RGB of NGC 2257 closely resembles that of NGC 5897. The RGB parameters and the derived metallicity computed in Section 3.2 are in quite good agreement with those found in NGC 5897 by FFB: $(B - V)_{0,g} = 0.74$, $S = 5.9$ and from which a metallicity of $[Fe/H] = -1.75$ has been derived. In Figure 8 the mean ridge lines of NGC 5897 have been overplotted on the diagram of Figure 2. A shift of $\Delta(B - V) = -0.06$ in colour and of $\Delta V = 2.75$ in magnitude have been applied to NGC 5897 ridge lines in order to take into account the different reddening and distance modulus (since for NGC 5897 $E(B - V) = 0.10$ and $V_{HB} = 16.35$). The NGC 5897 mean ridge line reproduces well the RGB in NGC 2257, except for the quoted very bright region where the RGB of 2257 flattens (see Section 3.1 and 3.2), supporting the "photometric" indication of a comparable metallicity.

Concerning the HB morphology, in both clusters HB stars mainly populate the blue side of the instability strip, even though the HB tends further blueward in NGC 5897. In particular, NGC 2257 shows $\sim 15\%$ of the HB stars on the red side of the instability strip and $\sim 23\%$ are RR Lyrae, in the case of NGC 5897 almost the whole HB population is located on the blue side with only a few RR Lyrae. We can use the LZ parameter defined in Section 3.4 in order to quantify the differences in the HB morphology of the two clusters. The $LZ$ parameter is $LZ = 0.87$ in NGC 5897 while in Section 3.4 we found $LZ = 0.45$ for NGC 2257. In the scenario proposed by Lee 1993, this difference could be interpreted in terms of a difference in age between the two clusters, NGC 2257 beeing $\sim$ 2Gyr younger than NGC 5897. This results is in agreement with Figure 9 by Walker (1992) in which the old LMC clusters appear in the mean to be $\sim$ 1 Gyr younger than the mean age of the GGCs.

**Figure 8:** Mean ridge lines of NGC 5897 shifted to match the CMD of NGC 2257. Shifts of $\Delta(B-V) = -0.06$ and $\Delta V = 2.75$ have been applied.

Another difference between the mean loci of the two clusters is visible in the TO and SGB region, though the photometry in NGC 5897 is not deep enough to reach the Main Sequence. From Figure 8 it can be seen that the mean ridge line of NGC 5897 does not reproduce the shapes of the SGB and the TO in NGC 2257, since the SGB shape looks "smoother" in NGC 5897 than in NGC 2257 and the TO of NGC 2257 seems to be brighter and bluer than that of NGC 5897, suggesting again that NGC 2257 is slightly younger than NGC 5897 (see the discussion on the age in Section 6).

# 6 THE AGE OF NGC 2257

In the previous section we obtained some preliminary clues which indicate that NGC 2257 is a few Gyrs younger than its twin-metallicity GGC NGC 5897. Since the TO point is well defined in our CMD, in this section we will try to get a quantitative estimate of the age of NGC 2257 appling the currently accreditated methods to date GGCs.

## 6.1 The $\Delta V_{TO}^{HB}$ method

The luminosity of the TO point is considered the "classical" clock of the stellar evolution. BCF used the TO luminosity referred to the HB level in order to obtain a new parameter, $\Delta V_{TO}^{HB} = V_{TO} - V_{HB}$, free by uncertainties on reddening and absolute calibrations.

In NGC 2257 the TO can be located at $V_{TO} = 22.40 \pm 0.10$, $(B-V)_{TO} = 0.35 \pm 0.05$. From this result and the figure for the HB we obtained in Section 3.2 we find $\Delta V_{TO}^{HB} = 3.30 \pm 0.15$, which is lower than the mean value $< 3.55 \pm 0.09 >$ obtained by BCF for a sample of 18 GGCs. This is a clear indication that the old LMC cluster NGC 2257 is younger than the average population of clusters in the Galaxy.

In order to have a quantitative estimate of the age of NGC 2257, with respect to a similar metallicity cluster in the Galaxy, we follow the prescriptions presented by Buonanno *et al* (1993), who proposed to use $\Delta V_{TO}^{HB}$ in a differential sense, (i.e. referring to another cluster, with comparable metallicity, chosen as reference). They defined the so-called vertical parameter as $\Delta = (\Delta_1 V_{TO}^{HB} - \Delta_2 V_{TO}^{HB})$ where $\Delta_1 V_{TO}^{HB}$ and $\Delta_2 V_{TO}^{HB}$ are the parameters for the programme and the reference cluster, respectively. They presented also a calibration of this differential parameter in terms of differences in age, using the VandeBerg and Bell models (1985):

$$\Delta log t_9 = (0.44 + 0.04[Fe/H])\Delta \quad (1)$$

where $t_9$ indicate the age expressed in Gyr.

**Figure 9:** The comparison with theoretical isochrones by Straniero and Chieffi 1991. The mean ridge line for NGC 2257 (small circles) is compared with two isochrones at 10 Gyrs (solid line) and 14 Gyrs (short dashed line).

An analogous "horizontal" parameter $\delta(B-V)$ has been proposed by VandenBerg, Bolte and Stetson (1990) and Sarajedini and Demarque (1990). This method uses the colour difference between the main sequence TO and the base of the giant branch as age-indicator (see also Chieffi and Straniero 1989). Since our photometry does not reach the MS with the accuracy needed to safely apply this method, we will base our analysis only on the parameter used by BCF and Buonanno et al 1993.

In the previous section we identified NGC 5897 as the metallicity twin of NGC 2257. Though the FFB's CMD of NGC 5897 does not reach the MS, the TO can be roughly located at $V_{TO} = 19.85 \pm 0.10$ giving $\Delta V_{TO}^{HB} = 3.50$, so that, for the cluster pair NGC 2257-NGC 5897 the $\Delta$ parameter is $\Delta = 3.30 - 3.50 = -0.20$. By using this value in the relation (1) we get $\Delta log t_9 = -0.074$ which means $\Delta t = -2.6 Gyr$ (for $t_9 = 15 Gyr$). On the other hand, another well studied GGC could be used as reference cluster having almost the same metallicity as NGC 2257: M 3 (Buonanno et al 1994a), for which $[Fe/H] = -1.66$. In this case $\Delta V_{TO}^{HB} = 3.52$, and the $\Delta$ parameter for the cluster pair NGC 2257-M 3 is $\Delta = 3.30 - 3.52 = -0.22$, giving $\Delta log t_9 = -0.08$, i.e.

$\Delta t = -2.8 Gyr$. Both these results indicate that NGC 2257 is about 2-3 Gyr younger than the mean clusters population with comparable metallicity in the Galaxy.

In the recent past, some GGCs have been found to be younger than the average globulars population in the Galaxy: Rup 106, Buonanno *et al* 1990, Arp 2 Buonanno *et al* 1994b, IC 4499 Ferraro *et al* 1994b (in preparation), Terzan 7 Buonanno *et al* 1994c, (in preparation). In particular Arp2 and IC 4499 have metallicity ( $[Fe/H] \sim -1.7$) comparable with that assumed for NGC 2257 and are thus very interesting reference galactic clusters to investigate the age of our "old" LMC cluster. Since Arp 2 has $\Delta V_{TO}^{HB} = 3.29$ the vertical parameter is $\Delta = 0.01$, corresponding to $\Delta log t_9 = 0.004$ and an age difference of 0.1 Gyr. In the case of IC 4499, we obtain $\Delta = 0.05$ (since $\Delta V_{TO}^{HB} = 3.25$) corresponding to $\Delta log t_9 = 0.019$ and an age difference of 0.6 Gyr.

These last two evaluations of the vertical parameter indicate that the ages of the oldest globulars in the LMC are similar to those of the peculiarly young GGC subsample recently discovered by Buonanno *et al* and Ferraro *et al*. This evidence suggests a possible connection between GGCs young subpopulation and the globulars in the LMC (Fusi Pecci *et al* 1994, in preparation).

## 6.2 The direct comparison to isochrones

More light on the metallicity and age determination in NGC 2257 can be done by direct comparison of the CMD to the isochrones.

*The metallicity question:*

As discussed in paragraph 3.3 the tip of the RGB shows a peculiarly low slope, in spite of all the available metallicity estimates which suggest that NGC 2257 is an intermediate metal poor cluster. This can be further investigated by comparing the CMD with available isochrones for old stellar populations.

Adopting the distance modulus from Section 4 one finds that only isochrones with [Fe/H] $\simeq$ -1.7 are able to reproduce the slope of the major portion of the red giant branch up to V $\simeq$ 17, supporting the adopted cluster metallicity.

*The reddening:*

In order to get a complete overlapping of the data with the isochrones we have to assume a a reddening even lower than the low one suggested by W89 $E(B - V) = 0.04$, so that our best fit is obtained assuming $E(B - V) = 0.0$. A value which would support the value found by W89 on the basis of RR Lyrae.

*The age:*

As discussed in the previous section there are some indications that NGC 2257 is younger than the mean GGC population. On these basis figure 9 compares the mean loci of NGC 2257 to two isochrones computed for [Fe/H] = - 1.7 and for the two alternative assumptions about the cluster age t = 10 and t = 16 billion year. One finds that the quoted distance and reddening drive theoretical prescriptions to nicely overlap observation, constraining in particular the age of the cluster within the already quoted limits and suggesting a value $t \sim 13 Gyr$.

## 7 CONCLUSIONS

The analysis performed throughout this paper allows us, for the first time, to reach a clear insight on the problem of the age of old MC clusters. NTT photometry going down

∼ 2 magnitudes below the TO point gives indeed rather firm constraints at least on the age of the cluster relative to Galactic Globulars, suggesting that this age should be safely estimated around 13 Gyrs (*i.e.* is about 2–3 Gyr younger than typical Galactic Globulars with comparable metallicity). Further investigations are obviously needed for other MC clusters to address some fundamental questions like whether old MC clusters are really coeval and, in any case, to compare the derived MC clusters' age with similar ages for the GGCs system.

## ACKNOWLEDGMENTS

We acknowledge Dr. A.R. Walker for helpul comment and suggestions. This work was partially supported by the Italian Ministry for the University and the Scientific Research (MURST). This paper is based on observations obtained at the European Southern Observatory, La Silla, Chile.

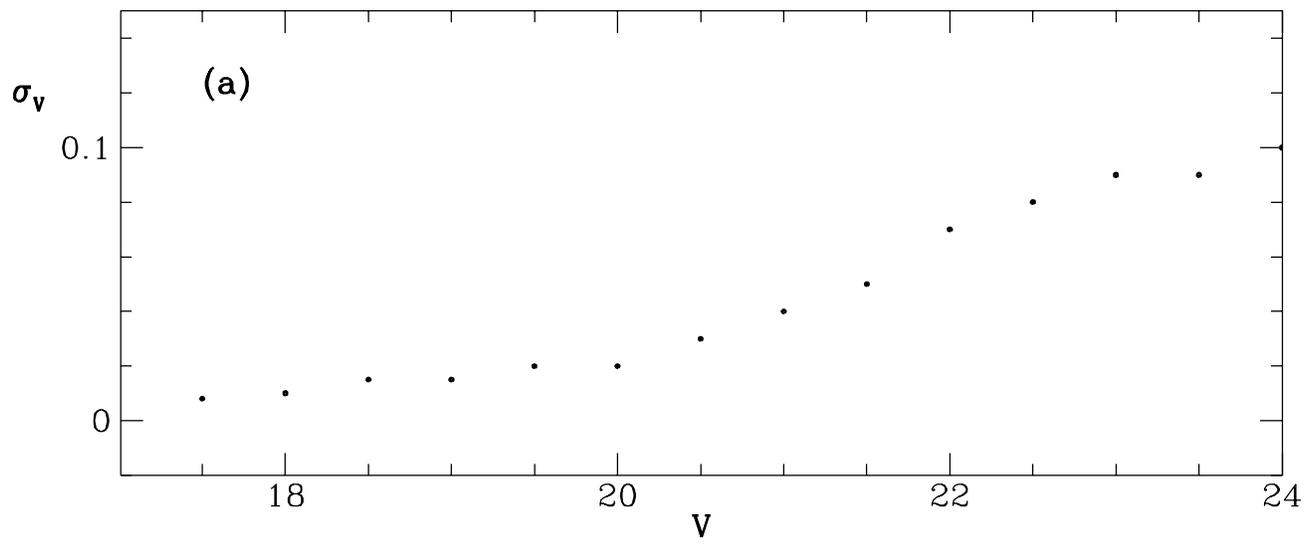
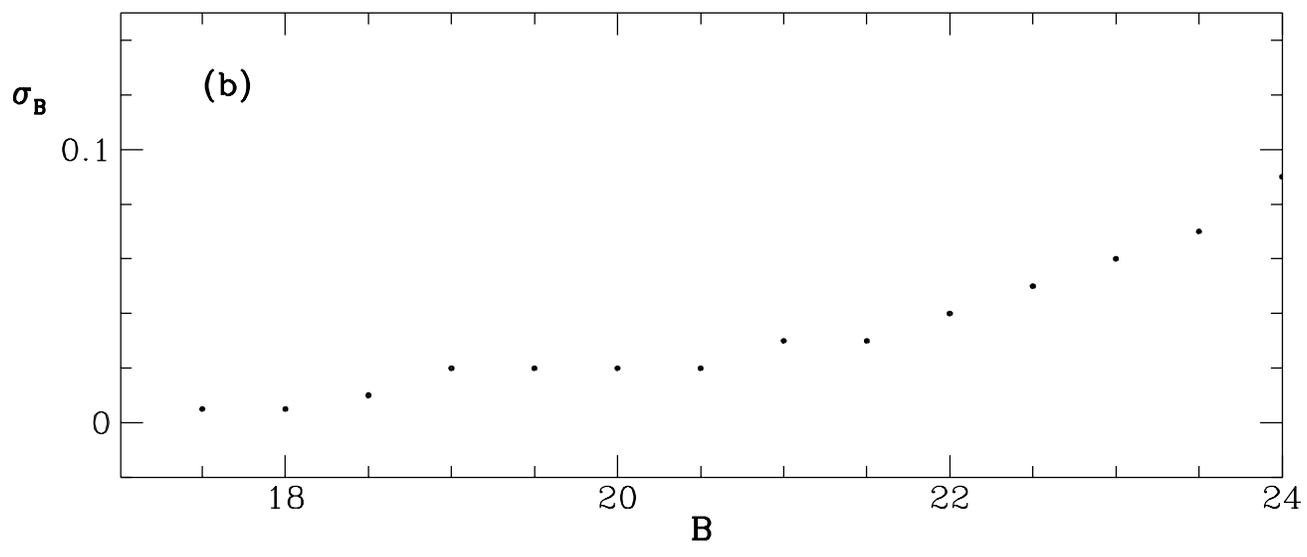

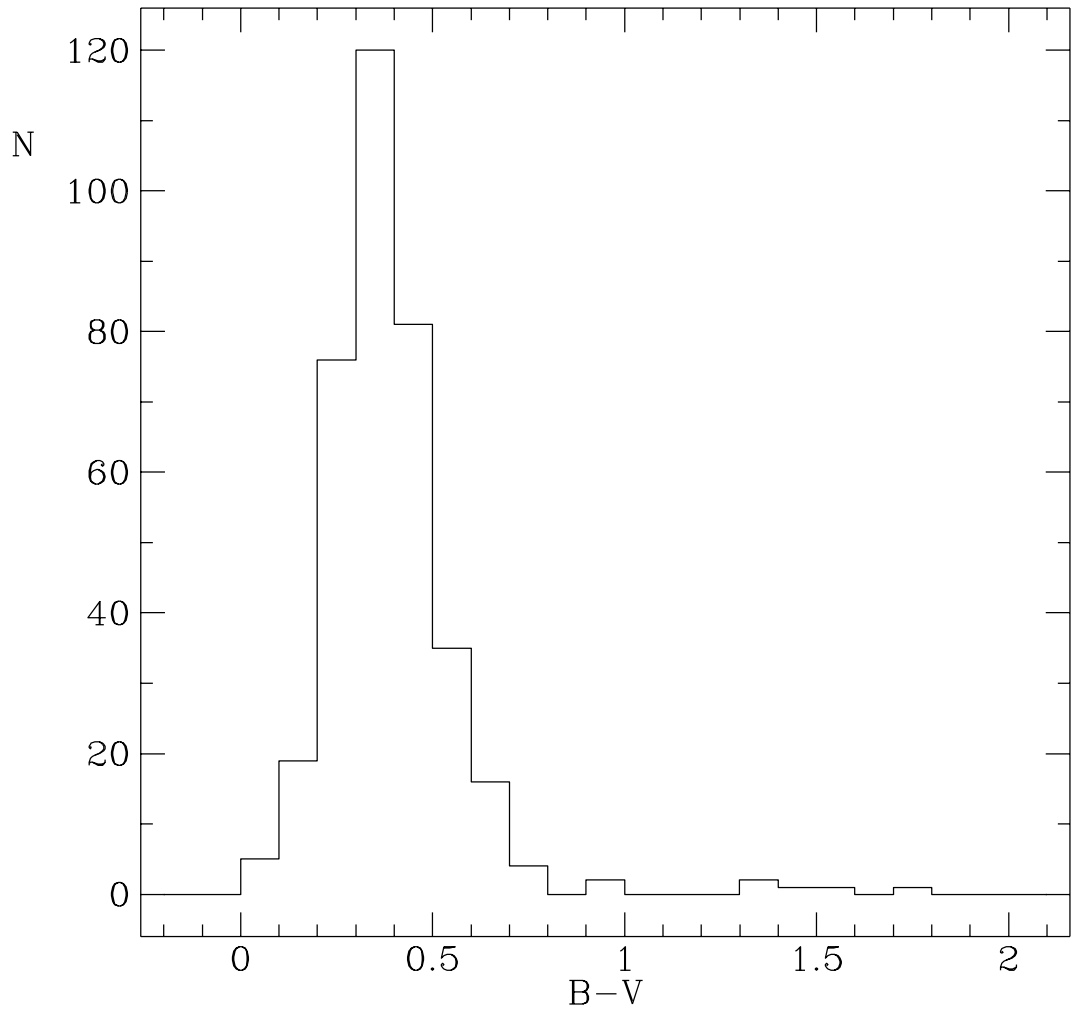

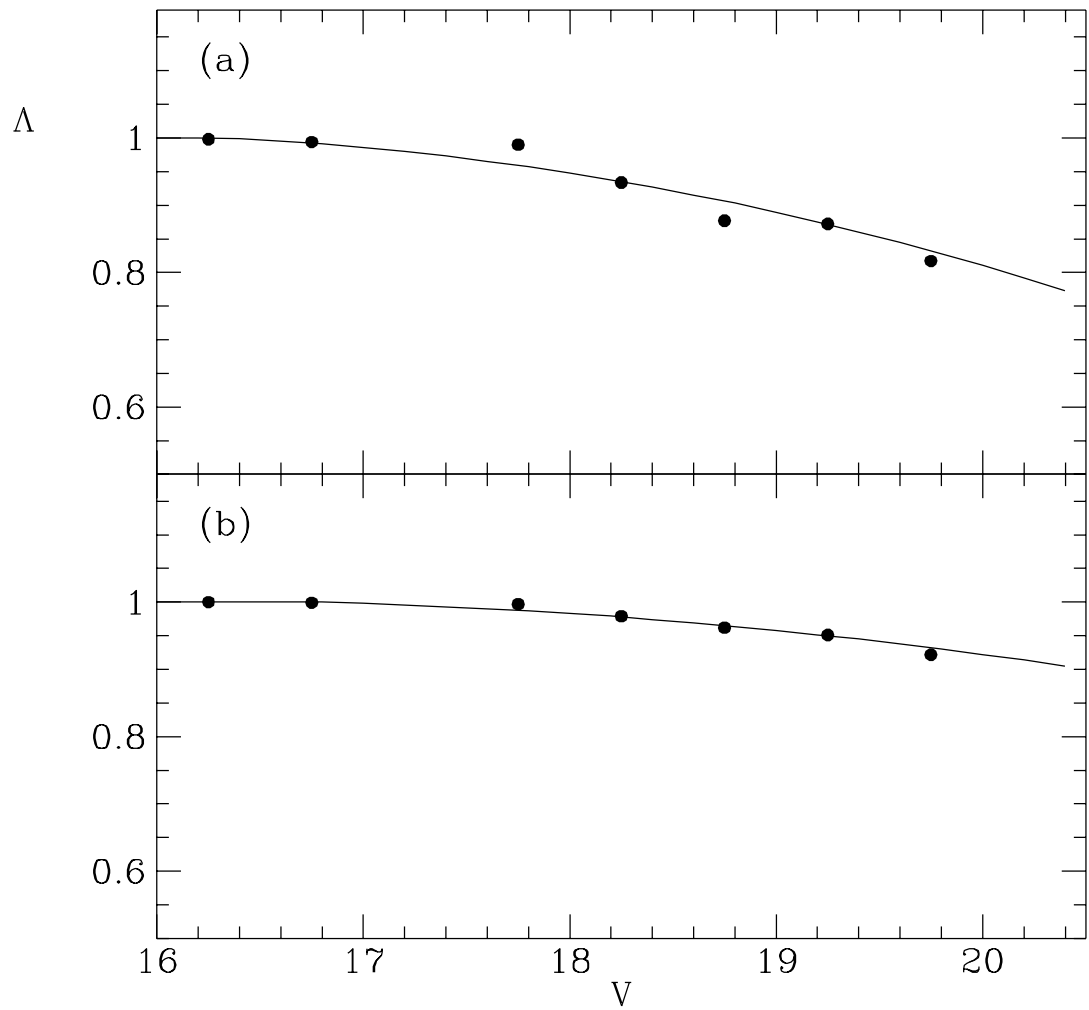

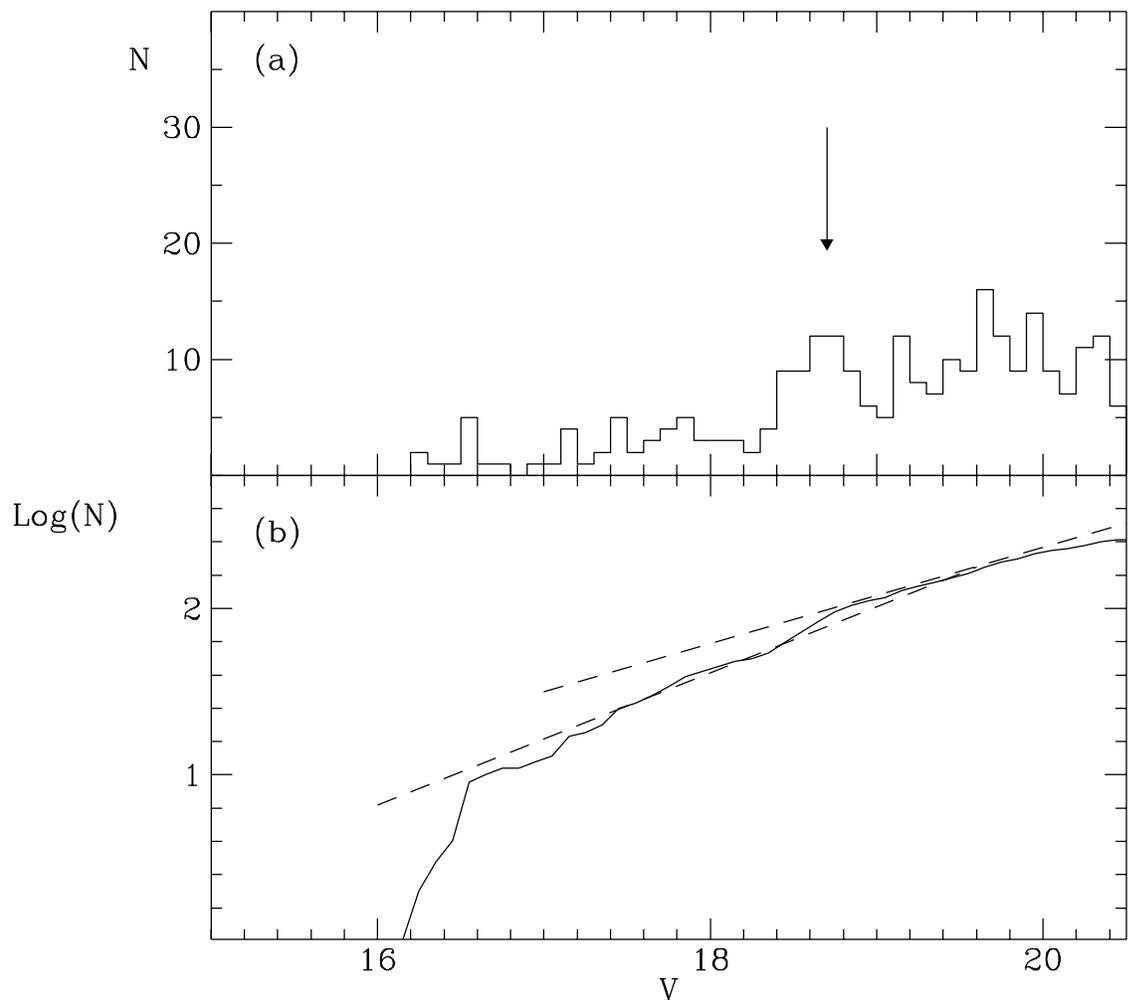

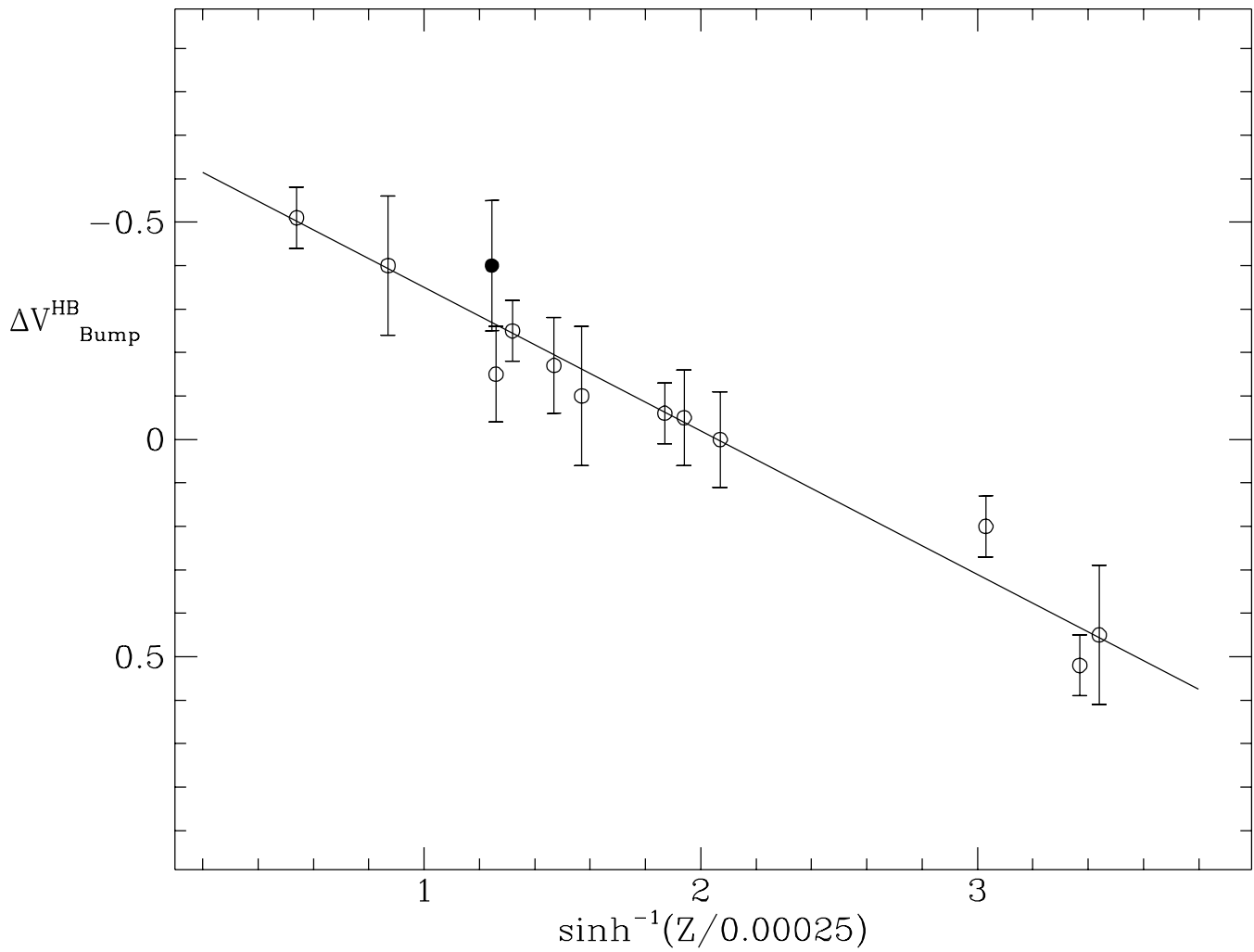

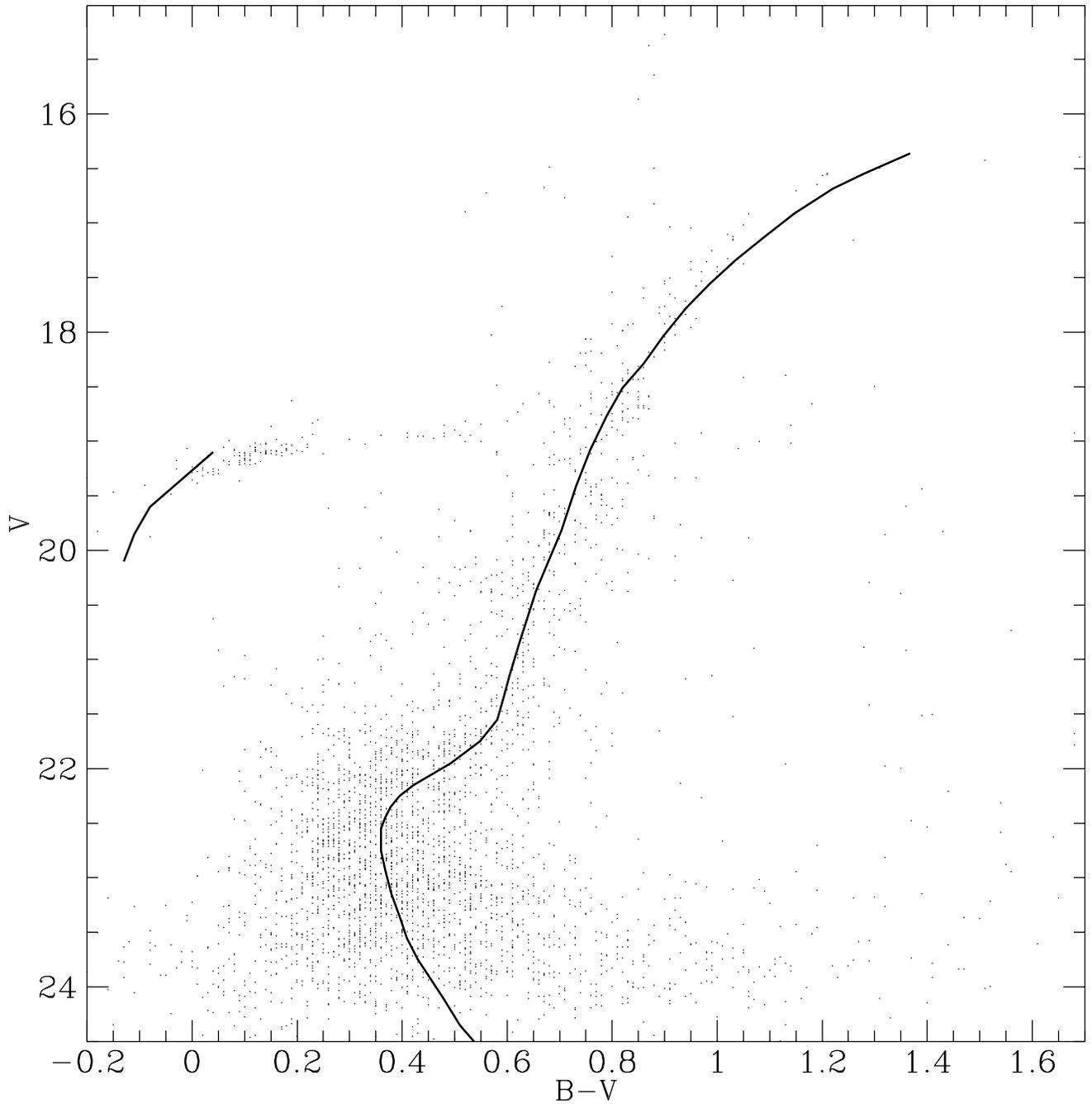

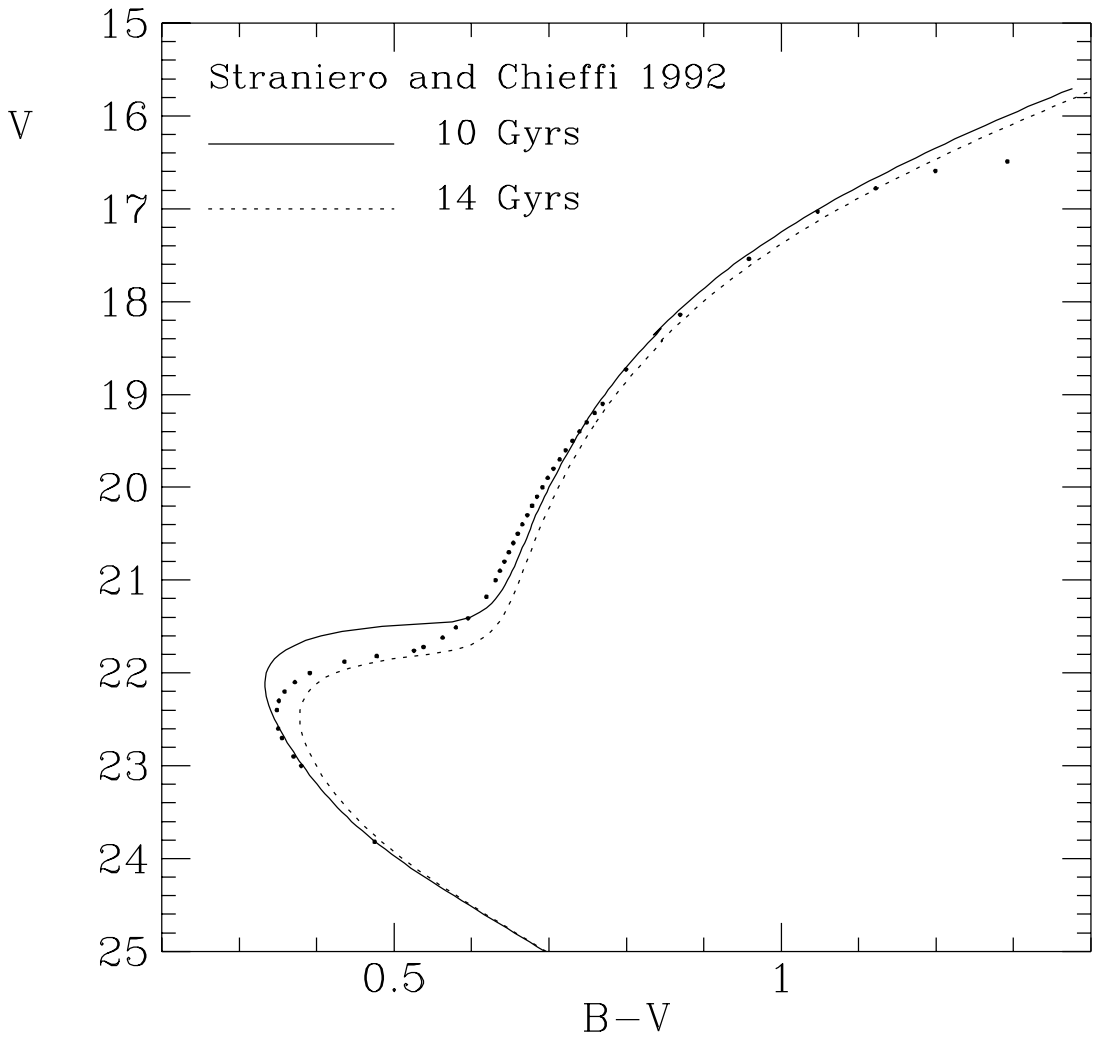